# What can we do when cosmology raises questions it cannot answer? Call in the philosophers?

**Joe Silk (Institut d'Astrophysique de Paris, the Johns Hopkins University, and Oxford University)**

MODERN cosmology has been spectacularly successful at explaining why the universe is as it is – a geometrically flat expanding space pockmarked with stars and galaxies. But this very success means that attempts to understand its origin increasingly stray into issues beyond physics and into the realm of philosophy, for which cosmologists rarely have any formal training. Likewise, when philosophers, untrained in astrophysical subtleties, pronounce on cosmology, the cosmologists are unimpressed. Clearly both groups have much to learn from each other

The philosophy of cosmology is not just unstructured pondering about where it all comes from or the meaning, if any, of our presence in the universe. It is the systematic survey of everything that possibly could have happened, and then to reconcile this with what actually did happen in our corner of physical reality.

What can be said, scientifically speaking, about these possibilities? For a start, many physicists are arriving at the conclusion that our visible universe is part of a larger space-time of infinite volume. Within this "multiverse", our universe would be but one of an infinity of space-time patches, each one outside the causal reach of any other. In infinite space, universes indistinguishable from ours would be repeated infinitely, as would every conceivable configuration of mass-energy permitted by the laws of physics. Moreover, the laws of physics themselves may even vary across the multiverse.

According to this scenario, it is difficult to avoid the conclusion that everything that can happen will happen, infinitely many times – an extraordinary proposition that takes us into philosophy's home territory.

The multiverse can be viewed as the continuation, or perhaps the climax, of a series of great shifts that dislodged the Earth, then the sun, and then our own galaxy from a special position at the centre of physical reality. Now even our visible universe – once hailed as being simply and inexplicably there, to paraphrase Bertrand Russell – is coming to be viewed as but one arbitrary patch of space-time within a manifold of infinite volume.

Support for this scenario stems from the "unnaturalness" of the fine-tuning that we observe in our surroundings. Consider the ratio of the mass of the neutron to that of the proton, about 1.001. It is exactly the ratio needed for nucleosynthesis – the creation of new atomic nuclei in stars. Without this process, we would not be here.

However, one fine-tuning issue stands out above all others: the cosmological constant, which accounts for the observed acceleration of the expansion of the universe. The predicted value, based on our understanding of the physics of the big bang, is larger than the observed value by a factor of $10^{120}$. The problem is, a cosmological constant much different from the observed value would make life as we know it impossible.

The increasingly popular solution to this conundrum is a model of the universe called eternal inflation. The idea is that inflation – the brief phase of ultrarapid expansion of the universe which occurred after the big bang spontaneously occurs again and again, budding off an infinity of new expanding universe domains. This scenario provides an explanation of the incredibly small value of our cosmological constant. If it has different values in each of the universe domains in the multiverse, then somewhere there has to be a universe with a small enough value to correspond to that observed in our universe.

How do we evaluate the odds of this happening? In a multiverse we would expect there to be relatively many universe domains with large values of the cosmological constant, but none of these allow gravitationally bound structures (such as our galaxy) to occur, so the likelihood of observing ourselves to be in one is essentially zero.

However, as the cosmological constant is decreased, we eventually reach a transition point where it becomes just small enough for gravitational structures to occur. Reduce it a bit further still, and you now get universes resembling ours. Given the increased likelihood of observing such a universe, the chances of our universe being one of these will be near its peak. Theoretical physicist Steven Weinberg used this reasoning to correctly predict the order of magnitude of the cosmological constant well before the acceleration of our universe was even measured.

Unfortunately this argument runs into conceptually murky water. The multiverse is infinite and it is not clear whether we can calculate the odds for anything to happen in an infinite volume of space-time. All we have is the single case of our apparently small but positive value of the cosmological constant, so it's hard to see how we could ever test whether or not Weinberg's prediction was a lucky coincidence. Such questions concerning infinity, and what one can reasonably infer from a single data point, are just the tip of the philosophical iceberg that cosmologists face.

Another conundrum is where the laws of physics come from. Even if these laws vary across the multiverse, there must be, so it seems, meta-laws that dictate the manner in which they are distributed. How can we, inhabitants on a planet in a solar system in a galaxy, meaningfully debate the origin of the laws of physics as well as the origins of something, the very universe, that we are part of? What about the parts of space-time we can never see? These regions could infinitely outnumber our visible patch. The laws of physics could differ there, for all we know.

We cannot settle any of these questions by experiment, and this is where philosophers enter the debate. Central to this is the so-called observational-selection effect, whereby an observation is influenced by the observer's "telescope", whatever form that may take. But what exactly is it to be an observer, or more specifically a "typical" observer, in a system where every possible sort of observer will come about infinitely many times? The same basic question, centred on the role of observers, is as fundamental to the science of the indefinitely large (cosmology) as it is to that of the infinitesimally small (quantum theory).

This key issue of typicality also confronted Austrian physicist and philosopher Ludwig Boltzmann. In 1897 he posited an infinite space-time as a means to explain how extraordinarily well-ordered the universe is compared with the state of high entropy (or disorder) predicted by thermodynamics. Given such an arena, where every conceivable combination of particle position and momenta would exist somewhere, he suggested that the orderliness around us might be that of an incredibly rare fluctuation within an infinite space-time.

But Boltzmann's reasoning was undermined by another, more absurd, conclusion. Rare fluctuations could also give rise to single momentary brains – self aware entities that spontaneously arises through random collisions of particles. Such "Boltzmann brains", the argument goes, are far more likely to arise than the entire visible universe or even the solar system. An infinity of space would therefore contain an infinitude of such disembodied brains, which would then be the "typical observer", not us.

Can this bizarre vision possibly be real, or does it indicate something fundamentally wrong with our notion of "typicality"? Or is our notion of "the observer" flawed – can thermodynamic fluctuations that give rise to Boltzmann's brains really suffice? Or could a futuristic supercomputer even play the Matrix-like role of a multitude of observers?

These questions about existence and our place in the universe are akin to those debated by philosophers throughout the ages. Now, for the first time, they are arising in concrete areas of scientific practice. Might a final, as-yet-undiscovered theory of quantum gravity reconcile all of these mysteries and, if so, could deep philosophical thinking pave the way, just as the work of philosophers such as David Hume and Ernst Mach did for Einstein?

Drawing the line between philosophy and physics has never been easy. Perhaps it is time to stop trying. The interface is ripe for exploration.